\documentclass[aps,prb,reprint,floatfix,twocolumn,citeautoscript,longbibliography]{revtex4-1}
\usepackage{graphicx}
\usepackage{ulem}
\usepackage{natbib}
\usepackage{xfrac}
\usepackage{bm,amsmath,amssymb,mathrsfs}
\usepackage[colorlinks=true, linkcolor=blue, citecolor=blue, urlcolor=blue, bookmarks=false]{hyperref} 
\usepackage[usenames]{xcolor}
\hypersetup{pdfborder=0 0 0,colorlinks=true,citecolor=blue,linkcolor=blue}

\def\ba{\begin{eqnarray}}
\def\ea{\end{eqnarray}}

\begin{document}
\title{Magnetic damping anisotropy in the two-dimensional van der Waals material Fe$_3$GeTe$_2$ from first principles}
\author{Pengtao Yang}
\author{Ruixi Liu}
\author{Zhe Yuan}
\affiliation{The Center for Advanced Quantum Studies and Department of Physics, Beijing Normal University, 100875 Beijing, China}
\author{Yi Liu}
\email{yiliu@bnu.edu.cn}
\affiliation{The Center for Advanced Quantum Studies and Department of Physics, Beijing Normal University, 100875 Beijing, China}
\date{\today}
\begin{abstract}
Magnetization relaxation in the two-dimensional itinerant ferromagnetic van der Waals material, Fe$_3$GeTe$_2$, below the Curie temperature is fundamentally important for applications to low-dimensional spintronics devices. We use first-principles scattering theory to calculate the temperature-dependent Gilbert damping for bulk and single-layer Fe$_3$GeTe$_2$. The calculated damping frequency of bulk Fe$_3$GeTe$_2$ increases monotonically with temperature because of the dominance of resistivitylike behavior. By contrast, a very weak temperature dependence is found for the damping frequency of a single layer, which is attributed to strong surface scattering in this highly confined geometry. A systematic study of the damping anisotropy reveals that orientational anisotropy is present in both bulk and single-layer Fe$_3$GeTe$_2$. Rotational anisotropy is significant at low temperatures for both the bulk and a single layer and is gradually diminished by temperature-induced disorder. The rotational anisotropy can be significantly enhanced by up to 430\% in gated single-layer Fe$_3$GeTe$_2$.\end{abstract}
\maketitle

\section{Introduction}
Newly emerged intrinsic two-dimensional (2D) ferromagnetic (FM) van der Waals (vdW) materials \cite{Gong:nat17,Huang:nat17,Fei:natm18,Deng:nat18,Wang:natn19,Otrokov:nat19} have become the subject of intense research. Weak vdW bonding facilitates the extraction of thin layers down to atomic thicknesses, whereas strong magnetocrystalline anisotropy protects long-range magnetic order. These materials provide an exciting arena to perform fundamental investigations on 2D magnetism and promising applications of low-dimensional spintronics devices. Among these materials, Fe$_3$GeTe$_2$ (FGT) is especially attractive for its itinerant ferromagnetism and metallicity, such that both spin and charge degrees of freedom can be exploited for designing functional devices. Bulk FGT has a relatively high Curie temperature ($T_\textrm{C}$) of approximately 220-230 K \cite{Deiseroth:ejic06,Chen:jpsj13,May:prb16,Liu:2dma17,Cai:apl20}. Atomically thin layers of FGT have lower $T_\textrm{C}$s, which, however, have been raised to room temperature (by ionic gating \cite{Deng:nat18}) and beyond (by patterning \cite{Li:nanol18}). As a FM metal at reasonably high temperature, FGT opens up vast opportunities for applications \cite{Gong:sc19,Wang:acsnano22,Li:nanos20,Yang:scad20,Ding:nanol19,Alghamdi:nanol19,Wang:scad19,Li:nanol19,Wang:nanol18,Yang:pccp20,Wang:acsnano20}.

The dynamical properties of FGT critically affect the applicability and performance of these proposed low-dimensional spintronics devices. The most salient of these properties is the dynamical dissipation of magnetization. It is usually described using a phenomenological parameter called Gilbert damping, which characterizes the efficiency of the instantaneous magnetization to align eventually with the effective magnetic field during its precessional motion. Although this parameter has been extensively studied in conventional FM materials, such as 3$d$ transition metals and alloys, two key issues with the Gilbert parameter of FGT remain to be addressed: the temperature dependence and anisotropy (one naturally expects anisotropic damping in FGT because of its layered structure and the strong magnetocrystalline anisotropy). Temperature-dependent Gilbert damping was first observed in Fe \cite{Heinrich:pssb66} and later more systematically in Fe, Co and Ni \cite{Bhagat:prb74,Heinrich:jap79,Khodadadi:prl20}.  A nonmonotonic temperature dependence has been found, for which a so-called ``conductivitylike'' component decreases with increasing temperature, usually at low temperatures, and a ``resistivitylike'' component increases with temperature, usually at high temperatures. This nonmonotonic behavior has been successfully described by the torque-correlation model \cite{Kambersky:cjp76} and reproduced by first-principles computations \cite{Gilmore:prl07,Kambersky:prb07,Liu:prb11,Ebert:prl11}. Anisotropic damping was first theoretically predicted in FM metals\cite{Gilmore:prb10} and in noncollinear magnetic textures \cite{Yuan:prl14}. With different orientation of the equilibrium magnetization with respect to the crystallographic axes, the damping parameter can be quantitatively different in general. This is referred to as the orientational anisotropy. Even for the same equilibrium magnetization orientation in a single crystalline lattice, the magnetization may precess instantaneously along different directions resulting in the so-called rotational anisotropy.\cite{Gilmore:prb10} The orientational anisotropy of damping has been observed in recent experiments on single-crystal FM alloys \cite{Li:prl19,Xia:prb21,Zhang:ami22}, but the underlying physical mechanism remains unclear. 

The dimensionless Gilbert damping parameter $\alpha$ can be expressed in terms of a frequency $\lambda$ via $\lambda=\alpha\gamma M$, \cite{Heinrich:05} where $M=|\mathbf{M}|$ is the magnetization magnitude and $\gamma$ is the gyromagnetic ratio. Despite of the different dimensions, these two parameters are equivalent\cite{Stiles:prb07} and both present in literature for experimental \cite{Heinrich:pssb66,Bhagat:prb74,Heinrich:jap79,Khodadadi:prl20,Li:prl19,Xia:prb21,Zhang:ami22} and theoretical studies.\cite{Kambersky:cjp76,Gilmore:prl07,Gilmore:prb10,Liu:prb11,Yuan:prl14,Starikov:prb18,Liu:prl14,Zhao:prb18}

In this study, we systematically investigate temperature-dependent Gilbert damping in single-layer (SL) and bulk FGT using first-principles scattering theory. Considering that the magnetization perpendicular to the 2D atomic planes is favored by the strong magnetocrystalline anisotropy, we calculate the damping as a function of temperature below $T_\textrm{C}$ and find nearly temperature-independent damping in the SL and damping dominated by resistivitylike behavior in the bulk. Varying the equilibrium direction of the FGT magnetization produces a twofold symmetry in damping. When the magnetization is aligned inside the 2D planes, a remarkable rotational anisotropy in the Gilbert damping is present for in- and out-of-plane rotating magnetization.

This paper is organized as follows. The crystalline structure of SL and bulk FGT is briefly introduced in Sec. \ref{sec:method}, followed by a description of our theoretical methods and computational details. The calculated temperature-dependent damping in SL and bulk FGT is presented in Sec. \ref{sec:temp}. The two types of damping anisotropy, i.e., orientational and rotational anisotropy, are analyzed in Sect. \ref{sec:anisotropy}. Conclusions are drawn in Sec. \ref{sec:conclusion}.
\section{Geometric structure of FGT and computational methods \label{sec:method}}
\begin{figure}[t]
\begin{center}
\includegraphics[width=0.9\columnwidth]{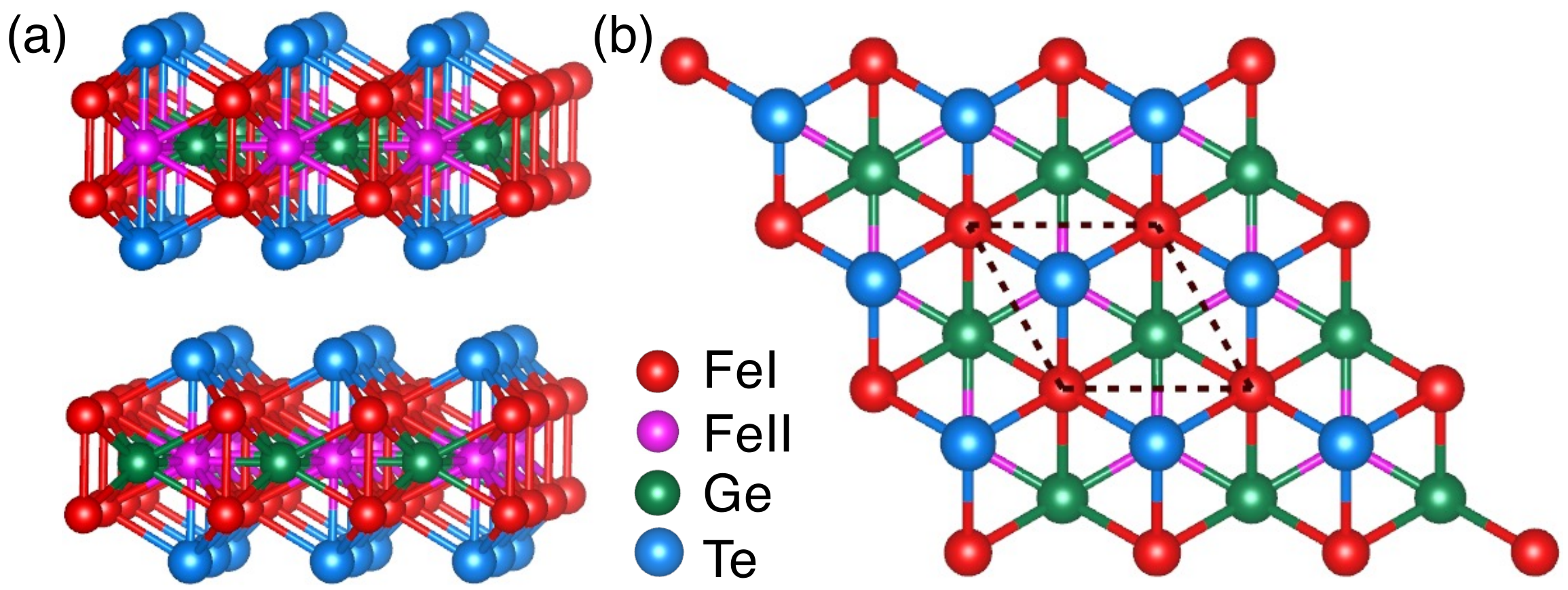}
\end{center}
\caption{(a) Side and (b) top view of the lattice structure for bulk Fe$_3$GeTe$_2$. The black dashed frame delineates the in-plane unit cell.}\label{fig:struct}
\end{figure}

The lattice structure of FGT is shown in Fig.~\ref{fig:struct}. Two different types of Fe atoms occupy inequivalent Wyckoff sites and are denoted as FeI and FeII. Five atomic layers stack along the $c$ axis to form an SL of FGT: Ge and FeII constitute the central atomic layer perpendicular to the $c$ axis, and two FeI layers and two Te layers are located symmetrically above and beneath the central layer, respectively. Single layers with ABAB$\dots$ stacking form the bulk FGT, where Layer A is translated in plane with respect to Layer B, such that the Ge atoms in Layer A lie on top of the Te and FeII atoms in Layer B.

The electronic structure of bulk and SL FGT has been determined using the linear augmented plane wave method \cite{Andersen:prb75} within the local density approximation (LDA). Different types of exchange-correlation functionals have been investigated in the literature, among which LDA was found to yield satisfactory structural and magnetic properties for FGT \cite{Zhuang:prb16}. We employ experimentally obtained lattice constants \cite{Deiseroth:ejic06} for bulk FGT calculations and obtain magnetic moments of 1.78~$\mu_\textrm{B}$ and 1.13~$\mu_\textrm{B}$ for the two types of Fe, respectively. The initial structure of a single layer is taken from the bulk lattice and fully relaxed, resulting in an in-plane constant $a=3.92$~{\AA}. A vacuum spacing of 11.76~{\AA} is chosen to exclude the interlayer interaction under periodic boundary conditions. The magnetic moments for the Fe atoms in SL FGT are obtained as 1.72~$\mu_\textrm{B}$ and 1.01~$\mu_\textrm{B}$. All the calculated magnetic moments are in good agreement with experimental \cite{Deiseroth:ejic06,Chen:jpsj13,Tian:prb19,Hu:ami20} and calculated values \cite{Zhuang:prb16,Shen:prb21,Jiang:prb22} reported in the literature.

The Gilbert damping calculation is performed using the scattering theory of magnetization dissipation proposed by Brataas {\it et al.}\cite{Brataas:prl08} Within this theory, a single domain FM metal is sandwiched between two nonmagnetic (NM) metal leads. The Gilbert damping that characterizes the energy dissipation during magnetization dynamics can be expressed in terms of a scattering matrix and its derivative with respect to the magnetization direction. We thus construct a two-terminal transport structure as Au$|$FGT$|$Au, where the Au lattice is slightly deformed to match that of FGT: we use 3$\times$1 and 4$\times$1 unit cells (UCs) of Au (001) to match the UCs of SL and bulk FGT, respectively. To investigate the effect of temperature on Gilbert damping, we use a frozen thermal lattice and spin disorder\cite{Liu:prb11,Liu:prb15,Starikov:prb18} to mimic lattice vibration and spin fluctuation at finite temperatures in FGT. The measured Debye temperature $\Theta_\textrm{D}=232$~K and temperature-dependent magnetization for the bulk \cite{Chen:jpsj13} and SL\cite{Fei:natm18} are employed to model the lattice and spin disorder. In the scattering calculations, lateral supercells are employed to satisfy periodic boundary conditions perpendicular to the transport direction. The electronic potentials required for the transport calculation are calculated self-consistently using a minimal basis of tight-binding linear muffin-tin orbitals (TB-LMTOs), and the resulting band structures for SL and bulk FGT effectively reproduce those obtained using the linear augmented plane wave method. Then, the scattering matrices consisting of reflection and transmission probability amplitudes for the Bloch wave functions incident from the NM leads are determined by the so-called ``wave function matching'' method, which is also implemented using TB-LMTOs \cite{Starikov:prb18}. Other computational details can be found in our previous publications \cite{Liu:prb11,Liu:prl14,Starikov:prb18,Zhao:prb18}. In this work, we focus on the damping with collective magnetization dynamics in the long-wave limit corresponding to the reported values in experiment via ferromagnetic resonance and time-resolved magneto-optical Kerr effect. The damping with a finite wavelength can be determined in our framework of scattering calculation \cite{Zhao:prb18} or using the torque-correlation model \cite{Gilmore:prb09}, but the wavelength dependence of damping is beyond the scope of the current study. 

\section{Temperature-dependent damping \label{sec:temp}}
\begin{figure}[t]
\begin{center}
\includegraphics[width=0.7\columnwidth]{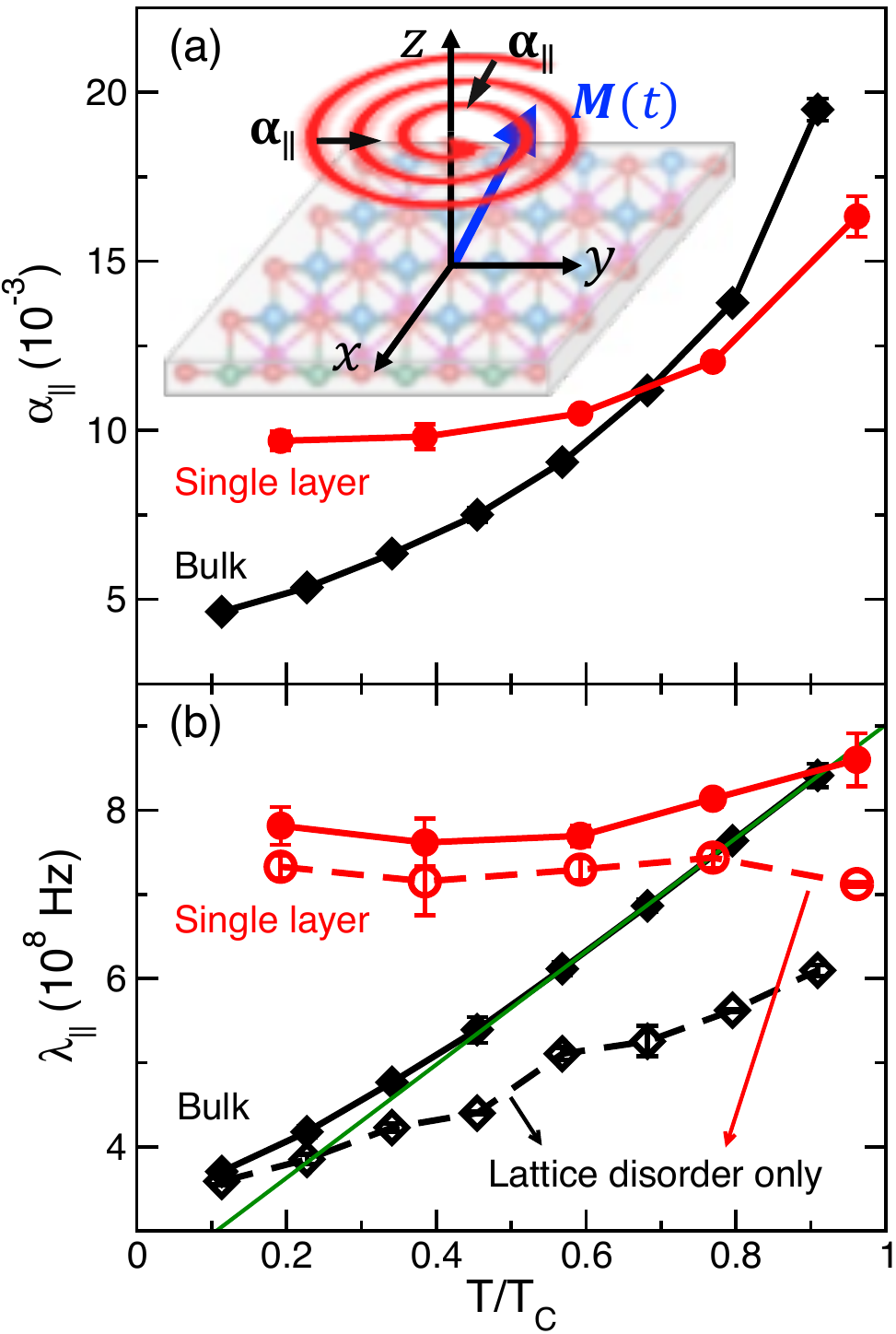}
\end{center}
\caption{The calculated dimensionless Gilbert damping parameter $\alpha_\|$ (a) and corresponding damping frequency $\lambda_\|$ (b) for single-layer and bulk Fe$_3$GeTe$_2$ as a function of temperature. The relaxation of the instantaneous magnetization $\mathbf{M}(t)$ results in a change in the in-plane magnetization component, which is parallel to the atomic planes, as schematized in the inset of (a). The empty symbols in (b) denote the damping frequencies that are calculated considering only thermal lattice disorder. The green line indicates the linear temperature dependence.}\label{fig:damping_T}
\end{figure}

The strong magnetocrystalline anisotropy of FGT results in the equilibrium magnetization being naturally perpendicular to the atomic layers. Slightly excited magnetization deviates from the plane normal (denoted as $\hat{z}$) and relaxes back by dissipating energy and angular momentum, as schematized in the inset of Fig.~\ref{fig:damping_T}(a). The Gilbert damping parameter $\alpha_\|$ describes the efficiency of such a dissipative process. The calculated $\alpha_\|$ of SL and bulk FGT is plotted in Fig.~\ref{fig:damping_T}(a) as a function of temperature. The damping for both increases monotonically with the temperature. This behavior resembles the so-called ``resistivitylike'' damping observed in many single-crystal FM metals \cite{Heinrich:pssb66,Bhagat:prb74,Heinrich:jap79}. However, the damping $\alpha_\|$ for the bulk tends to diverge as the temperature approaches $T_\textrm{C}$. This divergence originates from vanishing magnetization, as has been found in three-dimensional FM alloys \cite{Zhao:prb18}. Therefore, as temperatures approaching $T_\textrm{C}$, it is more appropriate to use the damping frequency parameter $\lambda=\alpha\gamma M$. 

The calculated damping frequencies are shown in Fig.~\ref{fig:damping_T}(b). The damping of a SL FGT, $\lambda^\textrm{S}_\|$, is larger than the damping of the bulk, $\lambda^\textrm{B}_\|$, especially at low temperatures. This difference can be attributed to the strong surface effect of highly confined SL FGT. The lowered symmetry at the surface significantly enhances spin-orbit coupling (SOC) \cite{Zhou:prb17}, which enables the dissipation of angular momentum from electronic spins to the orbital degree of freedom and then into the lattice reservoir. In addition, as the thickness of a single layer is considerably smaller than the electronic mean free path, conduction electrons in FGT are strongly scattered by the surface. Therefore, the two necessary ingredients for Gilbert damping, namely, SOC and electronic scattering, are both enhanced in the SL compared with the bulk, resulting in a larger damping for the SL.

The calculated damping frequency $\lambda^\textrm{S}_\|$ remains nearly constant with increasing temperature, except for a minor increase at $T>0.6T_{\rm C}$. To gain further insight into the temperature effect, we perform the damping calculation considering only lattice disorder, where the calculated $\lambda_\textrm{lat}^\textrm{S}$ are plotted as red empty circles in Fig.~\ref{fig:damping_T}(b). Lattice-disorder-induced damping in the SL FGT, $\lambda_\textrm{lat}^\textrm{S}$, exhibits a very weak temperature dependence, indicating that increasing lattice vibration does not influence the damping frequency. The difference between $\lambda_\textrm{lat}^\textrm{S}$ and $\lambda^\textrm{S}_\|$ increases slightly only near $T_{\rm C}$, which can be attributed to the strong spin fluctuation. The overall weak temperature dependence in the damping for a single layer indicates that a non-thermal disorder scattering mechanism is dominant: the strong surface scattering in such a thin layer (only a few {\AA}) combined with the enhanced SOC at the surfaces is the main channel for the magnetic damping in the SL FGT instead of spin fluctuation and lattice vibration. Gilbert damping with a similarly weak temperature dependence has also been found in a permalloy~\cite{Zhao:scr16, Starikov:prb18}, where chemical disorder scattering overwhelms thermally induced disorder. 

The temperature dependence of the bulk damping frequency is significantly different from that of the SL. The calculated bulk damping, $\lambda^\textrm{B}_\|$, (shown by the black solid diamonds in Fig.~\ref{fig:damping_T}(b)) increases linearly with the temperature. This typical resistivitylike behavior suggests that the interband transition in bulk FGT is the dominant damping mechanism  \cite{Gilmore:jap08}. We also calculate the damping frequency $\lambda_\textrm{lat}^\textrm{B}$ considering only lattice disorder, as shown as the black empty diamonds in Fig.~\ref{fig:damping_T}(b). Comparing the results corresponding to the solid and empty diamonds leads us to conclude that both lattice and spin disorder substantially contribute to damping in bulk FGT. As the temperature approaches $T_{\rm C}$, the bulk damping is comparable with that in the single layer.
\section{Anisotropic damping \label{sec:anisotropy}}
The damping torque exerted on the magnetization in the Landau-Lifshitz-Gilbert equation has the general form of $\mathbf M(t)\times[\bm{\tilde \alpha}\cdot\dot{\mathbf M}(t)]$, where the Gilbert damping parameter $\bm{\tilde \alpha}$ or the corresponding frequency is a tensor. This tensor and its elements depend on both the instantaneous $\mathbf M(t)$ and its time derivative $\dot{\mathbf M}(t)$, where the anisotropy has been extensively analyzed using theoretical models \cite{Steiauf:prb05} and first-principles calculations \cite{Gilmore:prb10,Yuan:prl14}. Following the definition given by Gilmore {\it et al.},\cite{Gilmore:prb10} we call the anisotropic damping that depends on the equilibrium orientation of $\mathbf M_{\rm eq}$ the orientational anisotropy and that depending on $\dot{\mathbf M}(t)$ the rotational anisotropy. Considering the layered structure of vdW materials, the lowered symmetry should result in remarkable anisotropy for the magnetization relaxation. Both the orientational and rotational anisotropy in bulk and SL FGT have been systematically analyzed in this section. Notably, the damping tensor is reduced to a scalar for the configuration shown in Fig.~\ref{fig:damping_T}.\begin{figure}[t]
\begin{center}
\includegraphics[width=0.8\columnwidth]{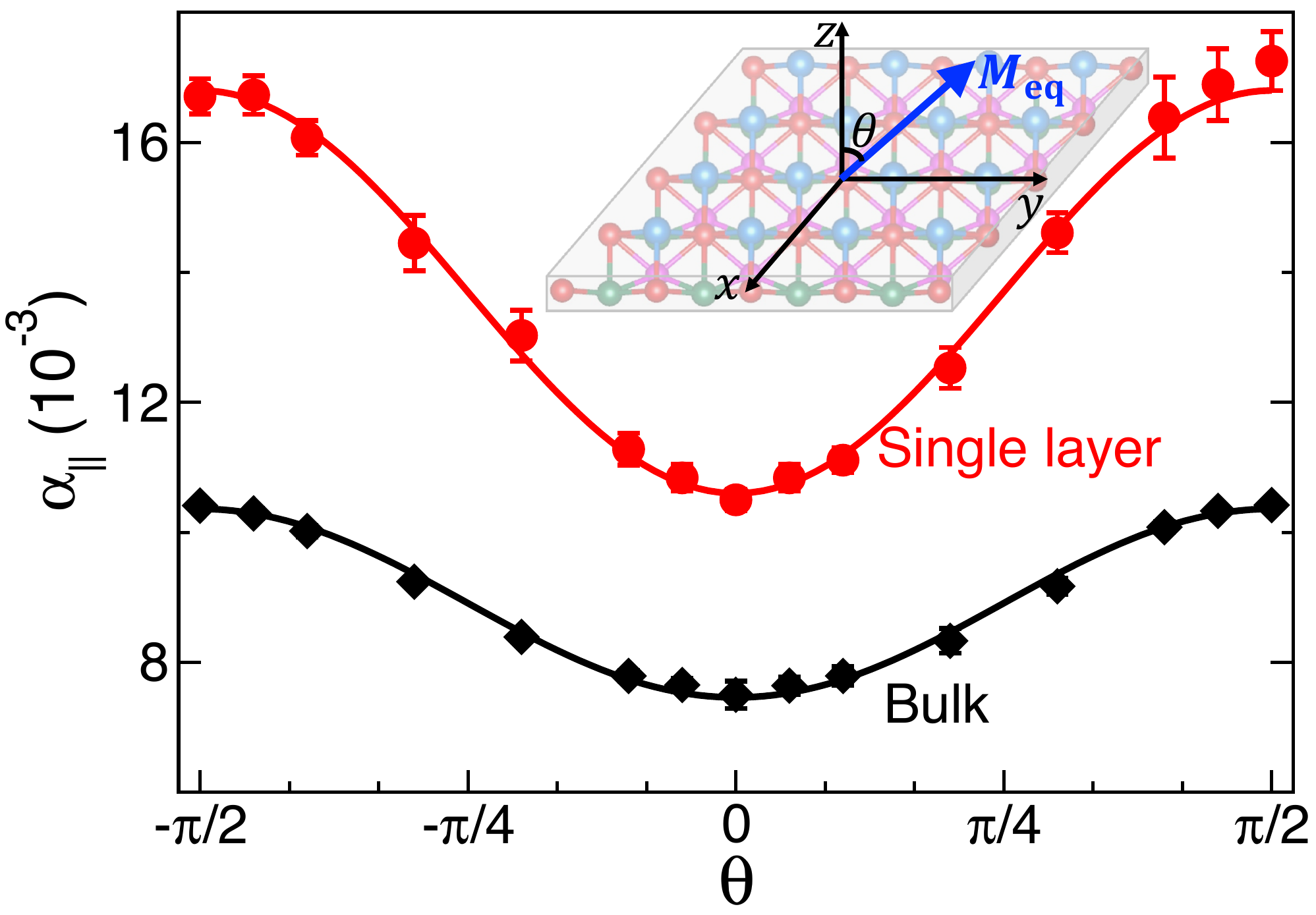}
\end{center}
\caption{The calculated Gilbert damping parameter $\alpha_\|$ for SL (red circles) and bulk FGT (black diamonds) as a function of the angle between the equilibrium magnetization $\mathbf M_\textrm{eq}$ and the atomic layer normal ($\hat{z}$) of Fe$_3$GeTe$_2$. The lines are fitted using $C_0+C_2\cos2\theta$.}\label{fig:theta}
\end{figure}

Under a large in-plane magnetic field, the perpendicular magnetization of FGT can be tilted toward the external field direction, which is defined as the $y$-axis without loss of generality. Thus, the angle between the equilibrium magnetization $\mathbf M_\textrm{eq}$ and the plane normal $\hat{z}$ is referred to as $\theta$, as shown in the inset of Fig.~\ref{fig:theta}. At $\theta=0$, as studied in Sec.~\ref{sec:temp}, $\alpha_{xx}=\alpha_{yy}=\alpha_\|$. For $\theta\ne0$, $\alpha_{xx}=\alpha_\|$ still holds, whereas the other diagonal element $\alpha_{yy}$ depends on specific values of $\theta$. Here, we focus on $\alpha_\|$ to study the orientational anisotropy of damping. The calculated in-plane damping $\alpha_\|$ is plotted as a function of $\theta$ in Fig.~\ref{fig:theta} for a SL at 77 K and bulk FGT at 100 K. The temperature is chosen in this way to obtain the same relative magnetization for the two systems, namely, $M/M_s=88\%$, according to the experimentally measured magnetization as a function of temperature \cite{Chen:jpsj13,Fei:natm18}. The same twofold symmetry is found for the damping parameters of both SL and bulk FGT, which can be effectively fitted using a $\cos2\theta$ term. As the magnetization rotates away from the easy axis, $\alpha_\|$ increases and reaches a maximum when the magnetization aligns inside the FGT layer. The changes, $[\alpha(\theta=\pm\pi/2)-\alpha(\theta=0)]/\alpha(\theta=0)$, are 62\% for the SL and 39\% for the bulk. A similar dependence of the damping on the magnetization orientation has been recently observed in single-crystal CoFe alloys\cite{Li:prl19,Xia:prb21}. The predicted anisotropic damping of FGT shown in Fig.~\ref{fig:theta} should analogously be experimentally observable.
\begin{figure}
\begin{center}
\includegraphics[width=0.75\columnwidth]{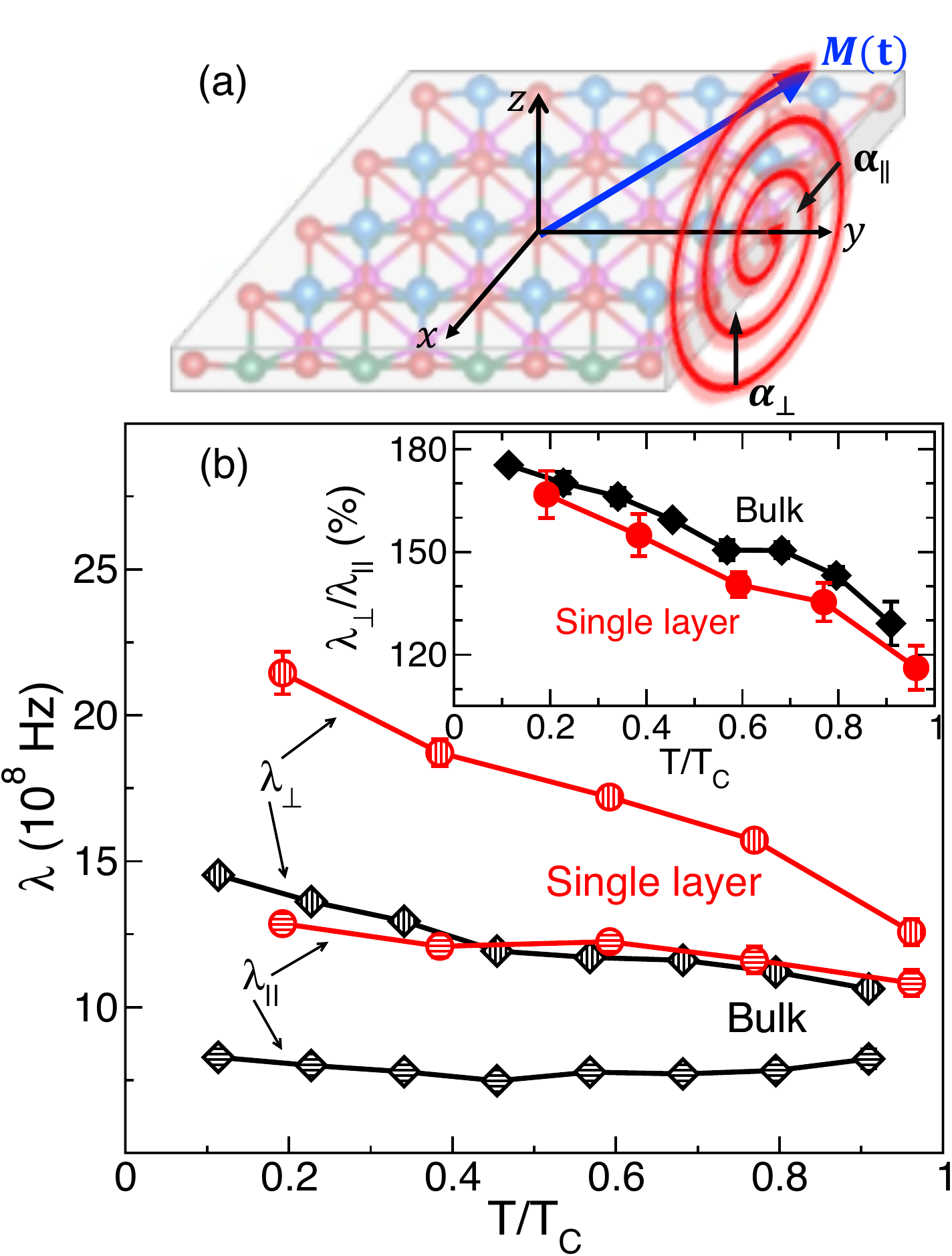}
\end{center}
\caption{(a) Schematic of damping with the equilibrium magnetization $\mathbf{M}_{\rm eq}$ lying inside the atomic plane. Then, the instantaneous magnetization $\mathbf M(t)$ dissipates both the in- and out-of-plane spin angular momentum. The two types of dissipation are denoted as $\alpha_\|$ ($\lambda_\|$) and $\alpha_{\perp}$ ($\lambda_{\perp}$). (b) The calculated Gilbert damping frequency $\lambda_{\|(\perp)}$ as a function of temperature for single-layer and bulk Fe$_3$GeTe$_2$. The inset shows the ratio of the two frequencies $\lambda_{\perp}/\lambda_\|$.}\label{fig:ani}
\end{figure}

The rotational anisotropy of damping \cite{Gilmore:prb10} in FGT is most significant when the equilibrium magnetization lies inside the atomic plane of FGT (along the hard axis), i.e., $\theta=\pm\pi/2$. As schematized in Fig.~\ref{fig:ani}(a), the magnetization $\mathbf M(t)$ loses its in- or out-of-plane components depending on the instantaneous precessional direction $\dot{\mathbf M}(t)$. In this case, one has $\alpha_{xx}=\alpha_\|$ and $\alpha_{zz}=\alpha_{\perp}$, whereas the off-diagonal elements of the damping tensor are guaranteed to remain zero by symmetry \cite{Starikov:prb18}. The calculated $\lambda_\|$ and $\lambda_\perp$ for SL and bulk FGT are shown as a function of temperature in Fig.~\ref{fig:ani}(b). For SL FGT, $\alpha_\|$ (as shown by the circles with horizontal hatching) is nearly independent of temperature, which is the same as for $\mathbf M_{\rm eq}$ along the easy axis. This result suggests that despite the sizable orientational anisotropy in the damping of SL FGT, the temperature has very little influence on the specific values of the damping frequency. The calculated $\lambda_\perp$ for the SL (shown by the red circles with vertical hatching) is considerably larger than $\lambda_\|$ at low temperatures but decreases with increasing temperature. $\lambda_\perp$ becomes comparable with $\lambda_\|$ near the Curie temperature, indicating that the rotational anisotropy is significantly diminished by temperature.

The calculated $\lambda_\|$ for bulk FGT with $\mathbf M_{\rm eq}$ along the hard axis (shown by the black diamonds with horizontal hatching) is temperature-independent, in sharp contrast to the linear temperature dependence of $\lambda_\|$ with $\mathbf M_{\rm eq}$ along the easy axis shown in Fig.~\ref{fig:damping_T}(b). This result suggests that the damping is already saturated in this case at a sufficiently large scattering rate, where saturated damping has also been found in FM Ni \cite{Bhagat:prb74}. The calculated $\lambda_{\perp}$ of bulk FGT is also larger than $\lambda_\|$ at low temperatures and slightly decreases with increasing temperature. We summarize the results for the rotationally anisotropic damping frequency by plotting the ratio between $\lambda_\perp$ and $\lambda_\|$ in the inset of Fig.~\ref{fig:ani}(b). The ratio for both SL and bulk FGT decreases with increasing temperature and approaches unity near $T_{\rm C}$. This behavior is consistent with the results calculated using the torque-correlation model \cite{Gilmore:prb10}, where rotationally anisotropic damping disappears gradually as the scattering rate increases. In highly disordered systems, the damping is more isotropic, as intuitively expected.

We emphasize that the calculated $\lambda_\perp$ values are distinct from those reported in previous studies in the literature \cite{Steiauf:prb05}, that is, $\lambda_\perp$ was found to vanish in single-crystal monoatomic FM layers based on the breathing Fermi surface model \cite{Kambersky:cjp70,Korenman:prb72,Kunes:prb02}. Interband scattering is neglected in the breathing Fermi surface model. However, the resistivitylike behavior of our calculated $\lambda_\|$ for bulk FGT shows that interband scattering plays an important role in this vdW FM material.
\begin{figure}[t]
\begin{center}
\includegraphics[width=0.75\columnwidth]{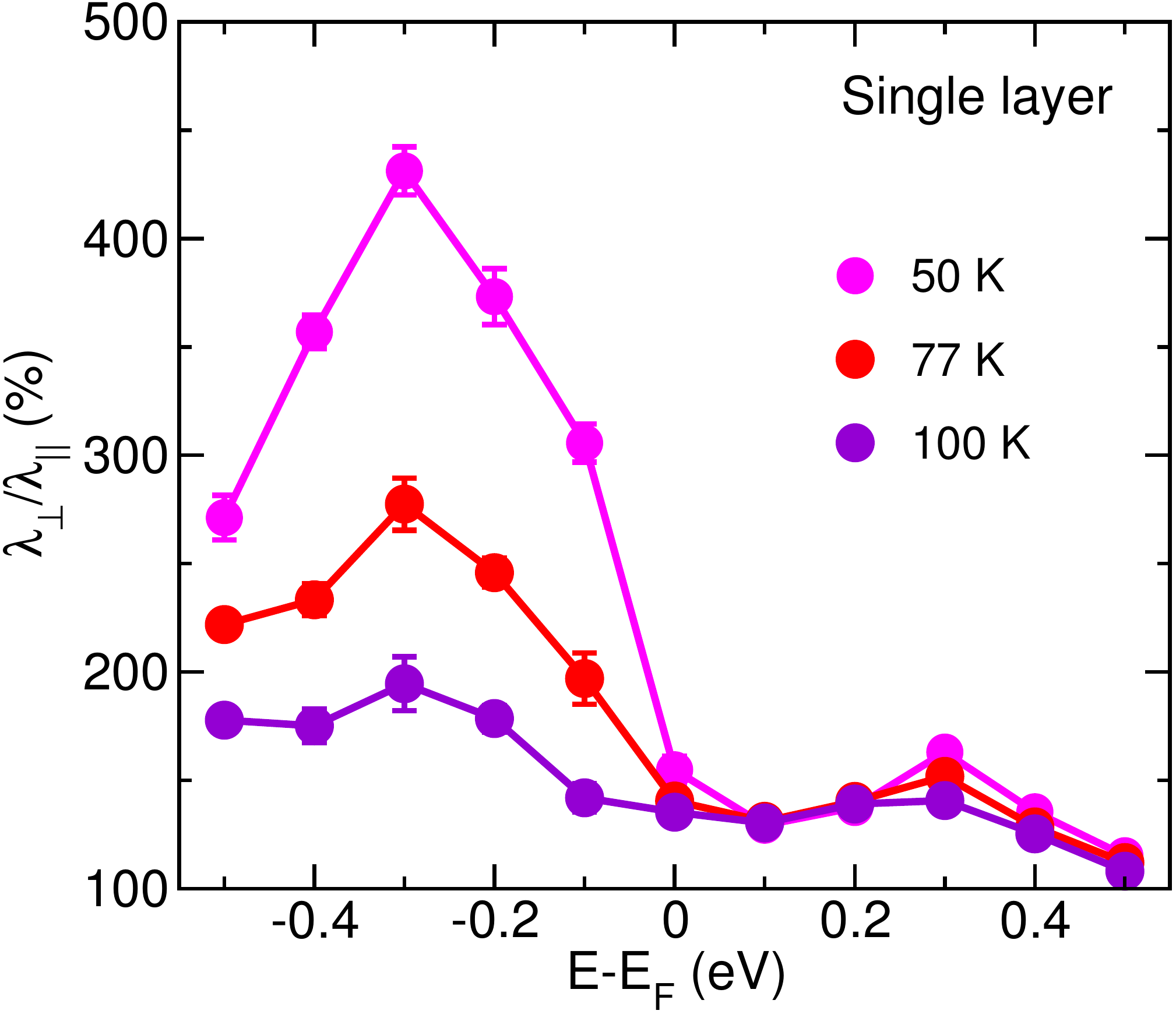}
\end{center}
\caption{The calculated rotational damping anisotropy for single-layer Fe$_3$GeTe$_2$ as a function of the Fermi energy at different temperatures. }\label{fig:ani_sl}
\end{figure}

One of the unique advantages of 2D vdW materials is the tunability of the electronic structure via electrical gating \cite{Deng:nat18,Zheng:prl20}. To simulate such a scenario, we slightly adjust the Fermi level $E_\textrm{F}$ of SL FGT without changing the band structure for simplicity. The calculated rotational anisotropy in the damping $\lambda_\perp/\lambda_\|$ of SL FGT is shown as a function of the Fermi energy in Fig.~\ref{fig:ani_sl}. At all the temperatures considered, the anisotropy ratio $\lambda_\perp/\lambda_\|$ increases dramatically as $E_\textrm{F}$ is lowered by 0.3~eV, especially at low temperatures, and only exhibits minor changes when $E_\textrm{F}$ is increased. At 50~K, the ratio $\lambda_\perp/\lambda_\|$ becomes as high as 430\%, which is almost three times larger than that obtained without gating. This result suggests that a small quantity of holes doped into SL FGT at low temperatures remarkably enhances the rotational damping anisotropy.

\section{Conclusions\label{sec:conclusion}}
We have systematically studied Gilbert damping in a 2D vdW FM material Fe$_3$GeTe$_2$ by using first-principles scattering calculations where the temperature-induced lattice vibration and spin fluctuation are modeled by frozen thermal lattice and spin disorder. When the magnetization is perpendicular to the 2D atomic plane, the damping frequency of bulk FGT increases linearly with the temperature, whereas that of SL FGT exhibits a weak temperature dependence. The difference can be attributed to surface scattering (which is absent in the bulk) dominating scattering due to temperature-induced disorder in SLs, which have a thickness smaller than the electronic mean free path. The anisotropy of Gilbert damping in this 2D vdW material has also been thoroughly investigated. The orientational anisotropy, which depends on the direction of the equilibrium magnetization with respect to the atomic planes, exhibits twofold rotational symmetry in both the bulk and SL. When the equilibrium magnetization is parallel to the atomic plane, the damping is significantly enhanced compared to that with the magnetization perpendicular to the atomic plane. The rotational anisotropic damping depending on the direction of motion of the instantaneous magnetization is remarkable with the equilibrium magnetization lying inside the atomic plane. With an out-of-plane component in the timederivative of the precessional magnetization, the damping frequency ($\lambda_{\perp}$) is much larger than the one where only in-plane magnetization is varying ($\lambda_{\|}$). The ratio $\lambda_{\perp}/\lambda_{\|}$ is larger than unity for both the bulk and a single layer and decreases with increasing temperature. In SL FGT, $\lambda_{\perp}/\lambda_{\|}$ can be enhanced up to 430\% by slight holedoping at 50 K.

Antiferromagnetic order has recently been discovered in 2D vdW materials (as reviewed in Ref. \onlinecite{Rahman:acsnano21} and the references therein) and some intriguing properties are found in their damping behaviors \cite{Afanasiev:scad21,Matthiesen:arxiv22}. Owing to the more complex magnetic order, more than a single parameter is necessary in describing the damping in antiferromagnetic dynamics.\cite{Liu:prm17,Yuan:epl19} It would be very interesting to study the magnetization relaxation in these 2D materials with more complex magnetic order. 

\acknowledgments
The authors are grateful to Professor Xiangang Wan at Nanjing University for his support and helpful discussions. Financial support for this study was provided by the National Natural Science Foundation of China (Grants No. 11734004 and No. 12174028).

\end{document}